\documentclass[%
reprint,
superscriptaddress,
bibnotes,
aps,
pra,
]{revtex4-2}

\usepackage{amsthm, amsmath, amssymb}
\usepackage{tensor, braket, physics}

\usepackage{graphicx, placeins}
\usepackage[caption=false]{subfig}

\usepackage{url}
\usepackage{comment}
\usepackage{xcolor}

\usepackage{hyperref, xcolor}
\usepackage[nameinlink,capitalize]{cleveref}
\hypersetup{
	colorlinks = true,
	linkbordercolor = {white},
	linkcolor={purple},
	citecolor={purple},
	urlcolor={blue}
}

\usepackage{cleveref}
\usepackage{orcidlink}

\begin{document}
\title{An analogical model for the stationary black holes by the flow field of space-time fluid around a 3-dimensional point sink}

\author{Behzad Ataei}
\email{behzad.ataei@alum.sharif.edu}
\affiliation{Department of Aerospace Engineering, Sharif University of Technology,Tehran, Iran}

\author{Ali Ayatollah Rafsanjani\,\orcidlink{0000-0002-3717-718X}}
 \email{aliayat@physics.sharif.edu}
\affiliation{Department of Physics, Sharif University of Technology,Tehran, Iran}
\affiliation{School of Physics, Institute for Research in Fundamental Sciences (IPM), Tehran, Iran}

\author{Alireza Bahrampour}
\affiliation{Department of Physics, Sharif University of Technology,Tehran, Iran}

\author{Mohammad Taeibi Rahni}
\affiliation{Department of Aerospace Engineering, Sharif University of Technology,Tehran, Iran}

\author{Mohammadreza Salimi}
\affiliation{Aerospace Research Institute, Sharif University of Technology, Tehran, Iran}

\author{Mehdi Golshani}
\affiliation{School of Physics, Institute for Research in Fundamental Sciences (IPM), Tehran, Iran}

\date{\today}

\begin{abstract}
Recent researches suggest an analogy between the theory of general relativity (GR) and fluid dynamics. As a result of this analogy, the Navier-Stokes equations and Einstein field equations are the same, and it is possible to study the properties of space-time by using fluid mechanics. In this paper, we present a new model to describe gravitational phenomena by an inviscid and compressible fluid called space-time fluid (STF). The analogy method is used to obtain the gravity field of both static and rotating masses from the flow field of STF around static and rotating point sinks. In addition, event horizons and the ergosphere of stationary black holes are defined based on our STF model. Then, we compare hydrodynamic forces exerted on a test particle with gravitational forces in the gravitoelectromagnetic approximation of the GR. As a natural consequence,  it is shown that inertial and gravitational masses are equivalent in this analogy. Finally, using the aspect of fluid dynamics, Mach's principle, weak equivalence principle, and information discontinuity on the event horizon are discussed.
\end{abstract}

\maketitle

\section{Introduction}
The analogy is a cognitive process that plays a significant role in problem-solving. It provides cross-fertilization of ideas between various research subjects by giving new ways of looking at problems. As examples of valuable analogies in physics, we can refer to the statistical theory of gases developed by Maxwell and Boltzmann, which established an analogy between gases and microscopic Newtonian rigid bodies through which they obtained entire laws of thermodynamics by those new microscopic degrees of freedom \citep{1klein1973development}. Another successful analogy was the gravitomagnetic hypothesis, which is an analogy between gravitation and electromagnetism that predicted some phenomenological aspects of general relativity (GR) before that \citep{2mashhoon2001gravitomagnetism}. The most well-known analogy of gravity in recent decades is the model of gravity and space-time by various condensed-matter systems, especially the fluid one \citep{11barcelo2011analogue}.

Fluid physics is a vibrant field that includes two essential scales, microscopic and macroscopic. Using the fluid-gravity analogies, we can extract microscopic and statistical attributes of space-time by the lattice-Boltzmann method and macroscopic one by Navier-Stokes equations. Therefore, these kinds of analogies help to understand the phenomenological effects of gravity better and obtain lessons that could be useful on the way to a quantum gravity theory.

Furthermore, due to the development of engineering tools in fluid dynamics, such analogies facilitate obtaining new solutions and simulations for critical aspects of gravity by using powerful instruments like computational fluid dynamics (CFD). Also, by establishing the fluid/gravity correspondence, we would have an exquisite attitude that could be an excellent source of intuition to solve the gravitational problems. For instance, there is no satisfactory solution for interior the Kerr metric, and we hope to model the gravitational field of a black hole by using the CFD to reach a clear picture of this problem. Moreover, it could be a good ground for experimental investigation of curved-space quantum field theory in condensed matter laboratories. It is possible to obtain a new entry for solving cosmological problems by using broad-range phenomena in fluid physics (like dark matter and energy).

In 1981 Unruh used a hydrodynamical analogy for a scalar field theory in an effective space-time using an acoustic wave in an irrotational inviscid barotropic fluid and then used this model to redescribe the black hole evaporation \citep{3brout1995hawking}. This first clue led us to look at gravitational phenomena, like Hawking's radiation, as emergent phenomena. After Unruh's primary model, many efforts were made to develop a fluid analogy to describe phenomena that correspond to fundamental curved-space quantum field theory, like black hole horizons, ergosphere, and Hawking's radiation, which has no direct relation with general relativity. Nowadays, these properties are known as attributes that emerge from unfamiliar microscopic degrees of freedom \citep{4unruh1981experimental}. 

In 1995, Jacobson derived Einstein's equation from the thermodynamical property of horizon and showed that one could look to Einstein's equation as an equation of state\citep{7jacobson1995thermodynamics}. Thereupon, one of the more explicit works that show we can look at gravity as an emergent force was 2011 Verlinde's paper which represented gravity as an entropic force \citep{8verlinde2011origin}. In the same year, Padmanabhan showed that Einstein's field equations are identical in form to Navier–Stokes equations of hydrodynamics \citep{9padmanabhan2011hydrodynamics}. His result led us to think that it is possible to obtain an identical description of large-scale gravitational phenomena by using the hydrodynamical analogy in addition to the statistical aspects of emergent gravity.

Also, in several articles, black holes were modeled by a two-dimensional sink or one-dimensional irrotational fluid dynamics in order to reproduce black hole thermodynamics \citep{10cadoni2004acoustic}, which in none of them, the flow field of space-time fluid (STF) and gravitational forces are described explicitly. Liberati and Visser have introduced the same idea and attempted to calculate the fluid flow field using the Schwarzschild metric. \citep{11barcelo2011analogue}

This article starts with time dilation in special relativity and compares it with gravitational time dilation phenomenologically. Then, using the classical inviscid barotropic fluid, the flow field of STF around a spherically symmetric mass distribution will be calculated. In this analogy, a gravitational mass is modeled by a point sink. We show that the flow field around a point sink introduces gravitational forces and the equivalence principle as a natural consequence. To validate our results, we will study all exerted forces on a test particle in STF flow to show that these forces are the same as the applied forces on probe B satellite by using the gravitomagnetic equations in the gravitoelectromagnetic approximation of the general relativity.

Afterward, in \cref{sec:blackholes}, the event horizon and ergosphere surface (ergosurface) of black holes are defined and calculated in static and rotational cases. We show that the sonic surfaces around the stationary and rotating point sinks correspond to the event horizon and ergosurface of stationary black holes. In \cref{sec:discussion}, we discuss the meaning of the Mach's principle, equivalence principle, preferred reference frame, and information discontinuity in both our fluid flow model and black holes.

\section{Time dilation}
Time dilation is the difference in elapsed time as measured by two observers. Here, we consider it phenomenologically as an effect observed in several experiments. It is either due to a relative velocity between observers (special relativistic time dilation) or a difference in gravitational potential between their locations (general relativistic gravitational time dilation)

\subsection{Relative and absolute time dilation}
According to the theory of Special relativity, the proportion of elapsed time rates for two observers at a speed of $\bold{v}$ relative to each other is
\begin{equation}
    \frac{d t^{\prime}}{d t}=\sqrt{1-\frac{\bold{v}^{2}}{c^{2}}},
    \label{eq:timedilation}
\end{equation}
Which \(d t\) is the rate of elapsed time measured by observer \(O\), and \(d t^{\prime}\) for observer \(O^{\prime}\). Notice that this equation is valid only in the reference frame of \(O\) and that \(O^{\prime}\) has another experience for the passage of time. This phenomenon has been confirmed in numerous experiments \citep{12mandelberg1962experimental,13hay1960measurement,14frisch1963measurement,15bailey1977measurements,16hafele1972around}

In addition to the time dilation between inertial observers, time passes slowly when an object is placed in a gravitational field. This phenomenon became validated in 1976 by gravity probe A of NASA. Unlike the relative velocity time dilation, observers agree on which clock is slower in the gravitational time dilation. We can also see this attribute in accelerated frames of reference. Einstein predicted the gravitational time dilation using these facts and assuming the equivalence principle.

The amount of time dilation for an object located at distant \(r\) from a static gravitational mass is
\begin{equation}
    \frac{d t^{\prime}}{d t}=\sqrt{1-\frac{2 G M}{r c^{2}}},
    \label{eq:Gtimedilation}
\end{equation}
with \(M\) the gravitational mass, \(G\) the gravitational constant, and \(\mathrm{c}\) the speed of light. Notice that \(d t\) corresponds to the observer \(O\) at infinity, which has no gravitational field, and \(d t^{\prime}\) is the proper time of observer \(O^{\prime}\) at position \(r\). Assume a rotating gravitational mass with specific angular momentum \(\bold{a}=\bold{J} / (M c)\) (with \(\bold{J}=I \bold{\Omega}\) the angular momentum); time dilation is written as
\begin{equation}
    \frac{d t^{\prime}}{d t}=\sqrt{1-\frac{2 G M r}{c^{2}\left(r^{2}+a^{2} \cos ^{2} \theta\right)}}.
    \label{eq:Rtimedilation}
\end{equation}

\subsection{STF flow field and time dilation}
First, we suppose STF as an inviscid fluid has filled the space. We propose that the relative time dilation is due to the relative velocity of the observer in the STF. So, it will be shown that by considering the gravitational time dilation as a phenomenological fact, we can obtain the equivalence principle and other gravitational phenomena. To reach this goal, first, we use \cref{eq:timedilation} and define the relative velocity between the STF flow and the object as \(\bold{u}_{s t f}\), and the speed of data transfer in STF as \(c_{s}\). So, we can write
\begin{equation}
    \frac{d t^{\prime}}{d t}=\sqrt{1-\frac{\bold{u}_{s t f}{ }^{2}}{c_{s}{ }^{2}}}.
    \label{eq:STFtimedilation}
\end{equation}

Here, \(O\) is the observer that is comoved with STF flow, and \(O^{\prime}\) has the relative velocity to the flow field. Due to the similarity of the expression \(\bold{u}_{s f t} / c_{s}\) to the Mach number in fluid dynamics, we called this expression the pseudo-Mach number as
\begin{equation}
   {\cal M} :=\frac{\left|\bold{u}_{s t f}\right|}{c_{s}}.
    \label{eq:pseudo-Mach}
\end{equation}

The pseudo-Mach number indicates the ratio of the velocity of STF flow to the speed of data transfer in STF.

Here, we assume the speed of signaling in the STF is equal to the speed of light, i.e., \(\left(c_{s}=c\right)\). So, the relation between the pseudo-Mach number and the time dilation is
\begin{equation}
    \frac{d t^{\prime}}{d t}=\sqrt{1-{\cal M}^{2}}.
    \label{eq:Machtimedilation}
\end{equation}

\section{STF flow field around a static and rotating mass}
\subsection{Static mass}
Consider \cref{eq:Machtimedilation} as the relation between the STF flow field and the gravitational time dilation. Using this, we can determine the STF flow field around a static point sink by an analogy between \cref{eq:Gtimedilation,eq:STFtimedilation}, as
\begin{equation}
    \bold{u}_{s t f}=-\sqrt{\frac{2 G M}{r}}\hat{r}.
    \label{eq:ustf}
\end{equation}

The spherical symmetry of the mass distribution led us to infer that the direction of the flow field has no \(\hat{\theta}\) and \(\hat{\phi}\) components and is just in the radial direction. Also, we demand an attraction force toward the source; thus, we assume an inward flow. This equation shows the radial flow field around a point sink. We can interpret it as a 3-dimensional point sink swallowing the STF.
\subsection{Rotating mass}
By considering a stationary rotating mass and comparing \cref{eq:Rtimedilation,eq:STFtimedilation}, the magnitude of the flow field will be given by
\begin{equation}
    \left|\bold{u}_{s t f}\right|=\sqrt{\frac{2 G M r}{r^{2}+a^{2} \cos ^{2} \theta}}.
    \label{eq:absustf}
\end{equation}
In this case, we cannot use the spherical symmetry to determine the velocity field components; however, axial symmetry implies that the field is independent of \(\hat{\theta}\). One can suggest that the radial component is the same as the static case and use the magnitude of the flow to calculate the \(\hat{\phi}\) component. Nevertheless, this is not true because in the poles \((\theta=0, \pi)\), the flow is just in the radial direction \footnote{Here, the Kerr metric and consequently the flow field of rotating masses have been written in the Boyer-Lindquist coordinates. Hence, the radial \(\hat{r}\) and transverse \(\hat{\phi}\) components are the confocal ellipses and hyperbolas in the ellipsoidal coordinate.} (at the rotational axis of the poles the flow has only the radial component). Hence,
\begin{equation}
    \bold{u}_{s t f}(\theta=0 ,\pi)=-\sqrt{\frac{2 G M r}{r^{2}+a^{2}}} \hat{r}.
    \label{eq:urstf}
\end{equation}

\section{static and rotating black holes\label{sec:blackholes}}
In our analogy, black holes are modeled as powerful point sinks toward which the STF flows, and at some points, the STF velocity reaches the speed of light. Thus, we will have sonic surfaces similar to normal fluids by considering the light same as the acoustic wave in STF. This section investigates the resulting sonic surfaces formed around the sinks. Correspondingly, the equivalence of the event horizon and the ergosurface around black holes with obtained sonic surfaces are being examined.

First of all, we investigate the compressibility of space-time fluid. For this purpose, the divergence of the flow field in \cref{eq:ustf} is calculated as
\begin{equation}
\nabla \cdot \bold{u}_{s t f} \neq 0.
\end{equation}
Consequently, we conclude that STF is compressible.

\subsection{static black hole}

In the STF model, the locus of the sonic surface around a static point sink corresponds to the event horizon of a stationary black hole in the Schwarzschild metric, which is  \citep{18rosen1971complete}:
\begin{equation}
    r_{s}=\frac{2 G M}{c^{2}}
\end{equation}

To obtain the sonic surface around a static point sink, we set the pseudo-Mach number in \cref{eq:pseudo-Mach} equal to one; 

\begin{eqnarray}
    {\cal M}&&=\frac{\left|\bold{u}_{s t f}\right|}{c_{s}}=\frac{\sqrt{\frac{2 G M}{r}}}{c_{s}}=1,
\end{eqnarray}
thus, the surface where the velocity of the STF is equal to the speed of light (sonic surface) is obtained as

\begin{eqnarray}
    r=\frac{2 G M}{c_{s}^{2}},
\end{eqnarray}
which is correspond to the Schwarzschild radius. 
\begin{figure}
    \centering
    \includegraphics[width = 8.6cm]{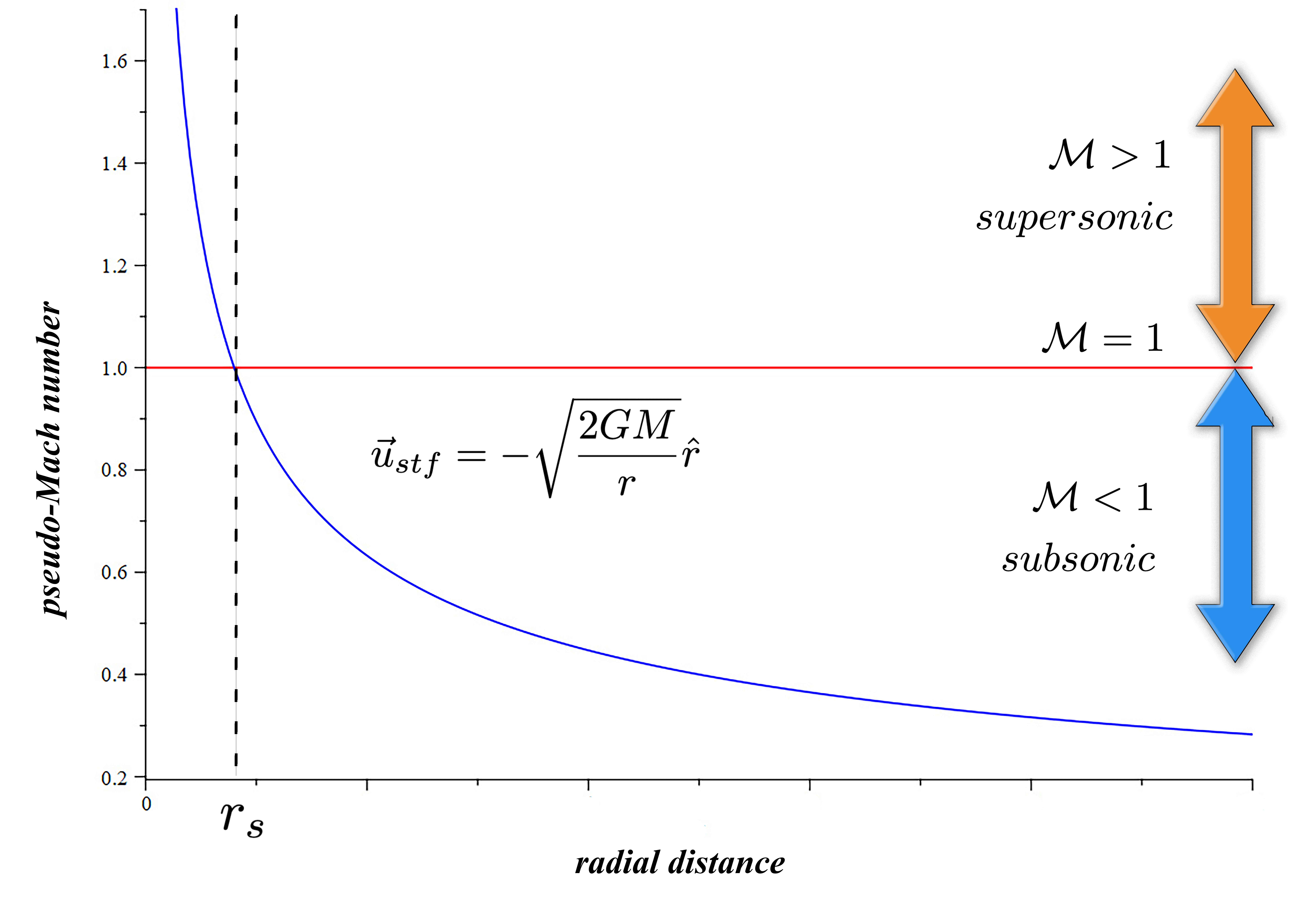}
    \caption{The STF radial velocity around static point sink}
    \label{fig:1}
\end{figure}

In \cref{fig:1}, the velocity of the STF around the sink is shown for a static point sink. In the radius \(r_{s}\), the velocity of the \(\mathrm{STF}\) is equal to the speed of light.

\subsection{rotating black hole}
\subsubsection{Event horizon and ergosphere in Kerr metric}
The event horizon and ergosurface around the rotating black hole are calculated using the Kerr metric.\citep{19weinberg1972gravitation} The radius of the event horizon
\begin{equation}
      r_{\pm}=\frac{r_{s} \pm \sqrt{r_{s}^{2}-4 a^{2}}}{2},
      \label{eq:horizons}
\end{equation}
and ergosurface are 
\begin{equation}
     r_{s_\pm}=\frac{r_{s} \pm \sqrt{r_{s}^{2}-4 a^{2} \cos ^{2} \theta}}{2}.
     \label{eq:ergos}
\end{equation}

\begin{figure}
    \centering
    \includegraphics[width = 8.6cm]{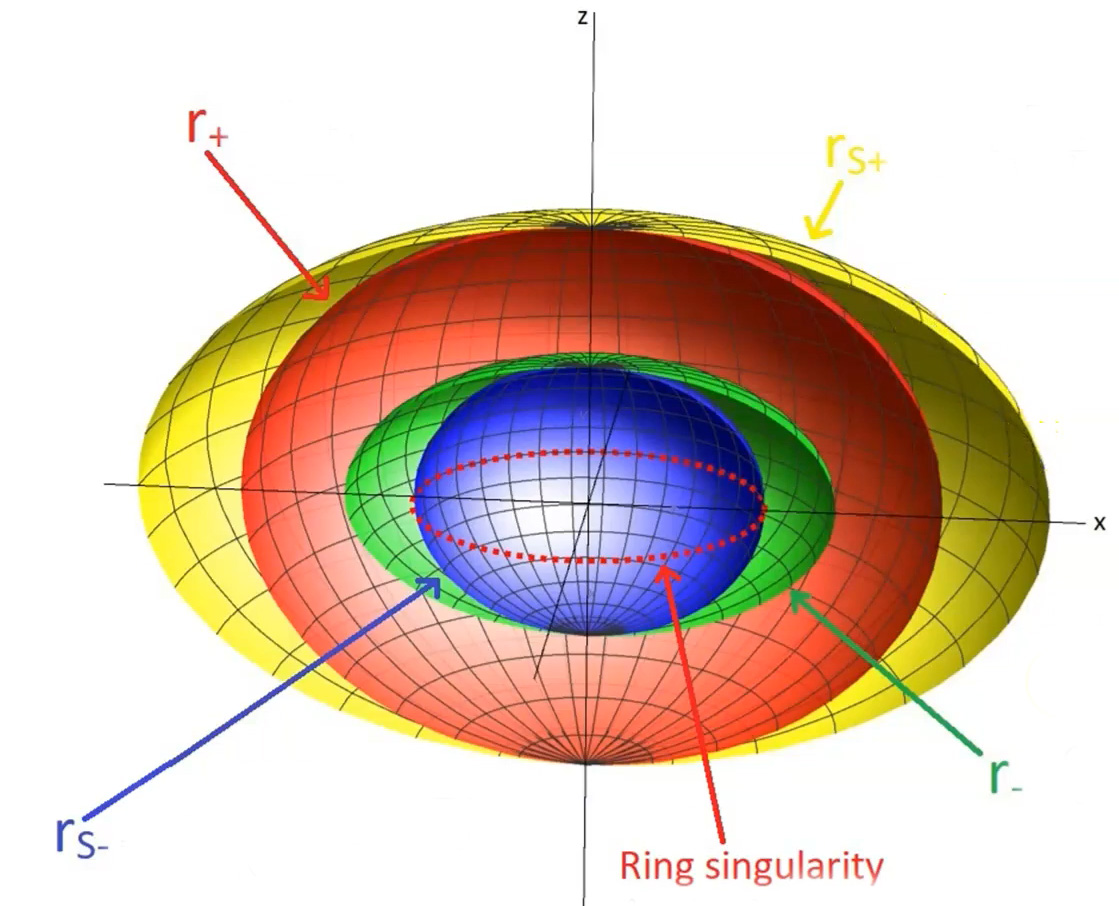}
    \caption{ergosphere and event horizons around the rotating black hole \citep{24youtube}}
    \label{fig:2}
\end{figure}

In \cref{fig:2}, we see the red surface as the outer event horizon with radius \(r_{+}\)and the green surface as the inner event horizon with radius \(r_{-}\), where these radii correspond to the values in \cref{eq:horizons}. In addition, the yellow surface is the outer ergosphere with radius \(r_{s_+}\), and the blue surface corresponds to the inner ergosphere with radius \(r_{s_-}\). These radii have corresponded to the values in \cref{eq:ergos}. 

\subsubsection{Sonic surface of a rotating point sink}
On the sonic surface around a rotating point sink, \(\bold{u}_{s t f}\) in \cref{eq:absustf,eq:urstf} must be equal to \(c_{s}\). We are setting \cref{eq:absustf}, the total velocity of the STF, equal to the speed of acoustic wave,

\begin{equation}
    {\cal M}=\frac{\left|\bold{u}_{s t f}\right|}{c_{s}}=\frac{\sqrt{\frac{2 G M r}{r^{2}+a^{2} \cos ^{2} \theta}}}{c_{s}}=1
    \label{eq:absmach=1}
\end{equation}
results in 
\begin{equation}
    r_{1,2}=\frac{r_{s} \pm \sqrt{r_{s}^{2}-4 a^{2} \cos ^{2} \theta}}{2}.
    \label{eq:pergos}
\end{equation}
As one can see, these surfaces are analogous to the ergosurface. An object within the \(r<r_{s_+}\) region cannot be stationary with respect to an outside observer at a far distance unless that object were to move faster than the speed of the acoustic wave concerning the local spacetime. However, objects can eject from there with a proper radial velocity.

To obtain the region where the test particles cannot eject from that, setting \cref{eq:urstf} (velocity in the direction of the radius of the STF around the rotating sink) equal to the speed of sound, we will have

\begin{equation}
   {\cal M}=\frac{\left|\bold{u}_{s t f}\right|}{c_{s}}=\frac{\sqrt{\frac{2 G M r}{r^{2}+a^{2}}}}{c_{s}}=1,
    \label{eq:rmach=1}
\end{equation}
So, points referring to areas where the radial velocity of the fluid becomes equal to the velocity of the sound wave are determined as
\begin{equation}
    r_{3,4}=\frac{r_{s} \pm \sqrt{r_{s}^{2}-4 a^{2}}}{2}.
    \label{eq:phorizon}
\end{equation}

\cref{eq:pergos,eq:phorizon} show surfaces where the total velocity and radial velocity of the STF flow are equal to the speed of light. In other words, the quasi-Mach number is equal to one at these surfaces, and information discontinuity occurs at the sonic surfaces in the fluid flow. Obtained surfaces correspond to the event horizon and the rotating black hole ergosurface.

According to the Euler equations, simplified forms of the Navier-Stokes equations, information travels through the pressure wave at the speed of sound in a compressible and inviscid flow. In subsonic regimes, the information can propagate both upstream and downstream of the flow; hence the equations have elliptical behavior. On the other hand, as velocity increases, information propagation becomes weaker until a singularity occurs at the speed of sound. In the meantime, the equations show parabolic behavior. This means that the information could no longer propagate upstream, resulting in information rupture. In the presented analogy, the sound speed in STF is equivalent to the speed of light propagation. The sound wave cannot pass through the sonic surface from downstream. In addition, there is a surface where the magnitude of the STF velocity is equal to the speed of information transfer in the fluid \citep{25anderson1990modern}.

Therefore, according to the \cref{eq:pergos}, if a sound wave moves in the radial direction, it can pass through the surface. In contrast, \cref{eq:phorizon} shows a surface where the STF velocity is equal to the speed of sound, and the surface prevents the sound wave from passing through even if it moves in a radial direction. This phenomenon leads to information disconnection. As mentioned before, several studies have shown an analogy between the sonic surface of the point sink and the event horizon of static black holes in two dimensions. Nevertheless, this article calculates sonic surfaces in three dimensions around static and rotating point sinks. Additionally, this study better explains the nature of information transfer in the event horizon and ergosphere. In the future, we will accurately model this subject with the help of CFD.

\section{Hydrodynamics and Gravitation}
Here we will study the phenomenological conformity between gravitational and hydrodynamic forces acting on a test particle in a two-phase inviscid fluid. Notice that the nature of dynamics in GR is due to the space-time geometry, and there is no concept of force in this theory. Thus, we should use the perturbation method to bring the two approaches on the same ground to have a fair comparison. Hence, utilizing the flow field obtained for the STF, we calculate the acting hydrodynamic forces on the test particle.

\subsection{Gravitoelectromagnetism}
Gravitoelectromagnetism, GEM for short, refers to a set of formal similarities between electromagnetic equations and relative gravity. In particular, between the Maxwell field equations and the valid approximation under certain conditions, to the Einstein field equations for GR. Gravitomagnetism is a widely used term that refers explicitly to the kinetic effects of gravity in comparison with the magnetic effects of a moving electric charge. The most common version of GEM is only valid away from separate sources and for slow-moving experimental particles. In this analogy, the GEM field is created around a rotating mass, the value of which is given by 

\begin{equation}
    \bold{B}_{g} =\frac{ G I}{2 c^{2} r^{3}}\left(\bold{\Omega}-{3(\bold{\Omega} \cdot \hat{r})\hat{r}}\right),
\end{equation}
and the force applied to the test particle is
\begin{equation}
    \bold{F} =m \bold{g}+4 m\left(\bold{v} \times \bold{B}_{g}\right),
    \label{eq:lorentzforce}
\end{equation}
which is like the Lorentz force \citep{26heaviside1893gravitational,27tisserand1890mouvement,28behera2018new,29mashhoon2003gravitoelectromagnetism,30mashhoon1984gravitational, 42thorne1988gravitomagnetism}. Here, \(\bold{B}_{g}\) is the gravitomagnetic field, \(\bold{v}\) is the velocity of the test particle, \(\bold{\Omega}\) is the angular velocity, and \(I\) is the area moment of inertia.

These approximated forces are due to the frame-dragging effect in GR. In 2011, NASA's Gravity Probe B experiment confirmed the forces of the gravitoelectromagnetic field. In this paper, the probe B satellite is considered as a spherical test particle orbiting a rotating planet \citep{31NASA}. Therefore we consider a spherical test particle, like the probe B satellite, orbiting a rotating planet.

\subsection{Hydrodynamic forces applied to a test particle}
The following forces are applied to a single spherical particle, with velocity \(\bold{v}\), in the inviscid flow field \(\bold{u}\) \citep{32tsuji2011particle,33maneshian2018bubble,34salari2013analytical,35putra2020modeling,36auton1987lift,37van2007drag}. By assuming the STF density to be equal to the particle density, \(\rho_{c}\), we have
\begin{align}
    &\bold{F}_{vm}+\bold{F}_{L}+\bold{F}_{p}=\left(\rho_{c} \forall_{p}\right) \frac{d \bold{v}}{d t},\label{eq:totalforce} 
\end{align}
with
\begin{align}
    &\bold{F}_{v m}=\frac{1}{2} \rho_{c} \forall_{p}\left(\frac{D \bold{u}}{D t}-\frac{d \bold{v}}{d t}\right), \\
    &\bold{F}_{p}=\rho_{c} \forall_{p}\left(\frac{D \bold{u}}{D t}\right),\\
    &\bold{F}_{L}=C_{L} \rho_{c} \forall_{p}(\bold{v} \times \bold{\omega}).
    \label{eq:vorticityforce}
\end{align}
\(\bold{F}_{v m}\) stands for the virtual mass, \(\bold{F}_{L}\) for the pressure gradient force, and \(\bold{F}_{p}\) for the lift force in these equations. The particle  fluid-mass is \(m_{p}=\rho_{c} \forall_{p}\), with \(\forall{ }_{p}\) the effective particle volume. Also, \(C_{L}\) is the lift force coefficient, and \(\bold{\omega}\) is the vorticity field. Given that the motion velocity of the test particle is \(\bold{v}\), the acceleration of the particle with respect to the comoving observer flows with STF is equal to:
\begin{equation}
    \bold{a}=\frac{d \bold{v}}{d t}
    \label{eq:acceleration}
\end{equation}

\subsubsection{Hydrodynamic forces due to the flow field around the static point sink:}
The radial acceleration field is calculated by the material derivative of the flow field in spherical coordinates, which for a steady flow is as follows \citep{38white2006viscous}:
\begin{equation}
    \frac{D {u}_{r}}{D t}={u}_{r} \frac{\partial {u}_{r}}{\partial r}+\frac{{u}_{\theta}}{r} \frac{\partial {u}_{r}}{\partial \theta}+\frac{{u}_{\phi}}{r \sin \theta} \frac{\partial {u}_{r}}{\partial \phi}
    \label{eq:materialderiv}
\end{equation}
By placing the velocity field of the static point sink, \cref{eq:ustf}, in \cref{eq:materialderiv}, we'll have
\begin{equation}
    \frac{D \bold{u}_{s t f}}{D t}=\frac{-G M}{r^{2}}\hat{r}.
    \label{eq:accfield}
\end{equation}

\cref{fig:3} shows the magnitude and direction of the STF acceleration field correspond to the gravity field. By placing \cref{eq:acceleration,eq:accfield} in \cref{eq:totalforce}, the forces acting on the test particle around the static point sink are acquired as follows (\(\rho_{\text {stf }}\) is STF density):

\begin{figure}
    \centering
    \includegraphics[width = 8.6cm]{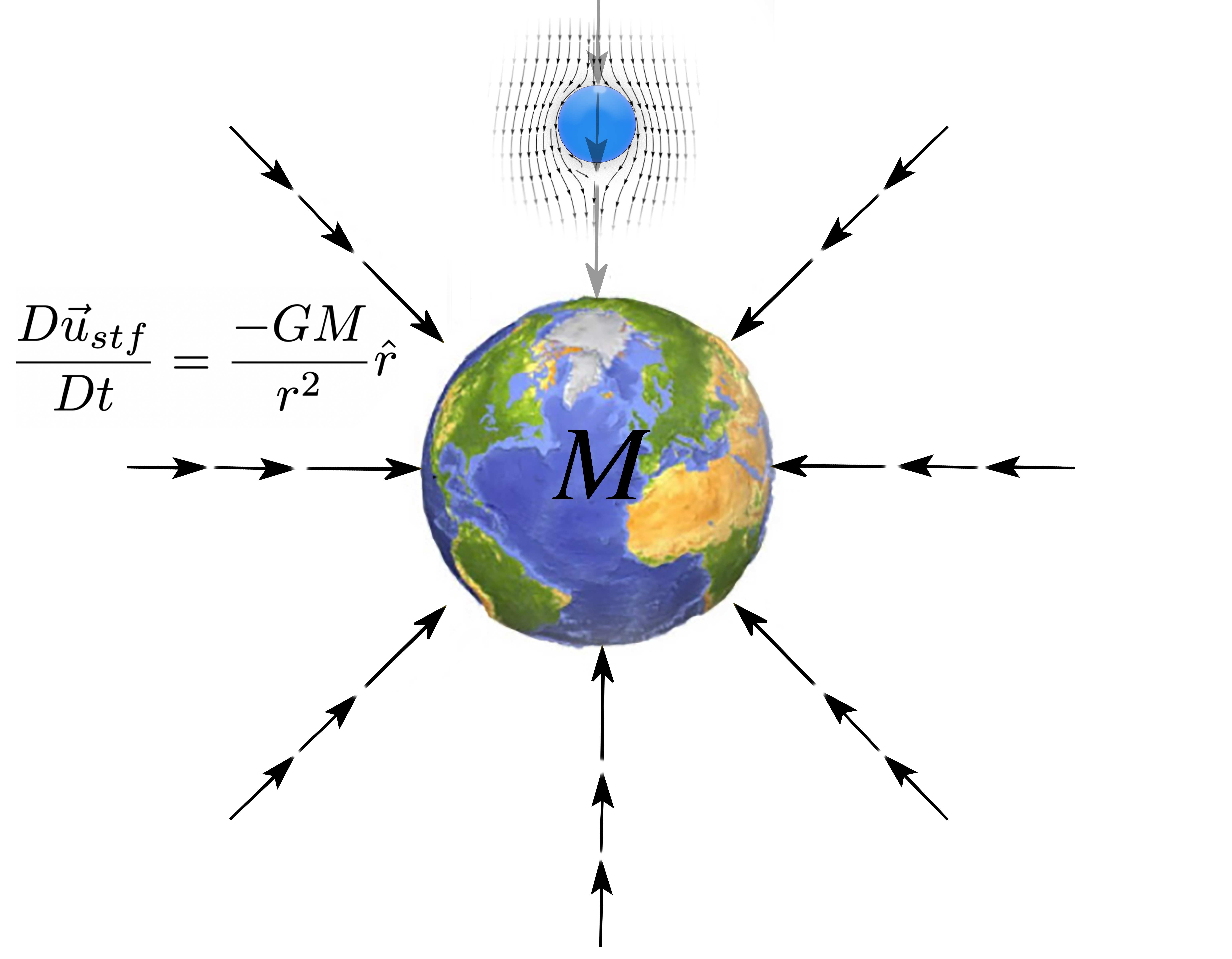}
    \caption{STF acceleration field around a point sink}
    \label{fig:3}
\end{figure}

\begin{eqnarray}
    \rho_{s t f} \forall_{p}\left(\frac{d \bold{v}}{d t}\right)&&=\rho_{s t f} \forall_{p}\left(\frac{-G M}{r^{2}}\right)\nonumber\\
&&+\frac{1}{2} \rho_{s t f} \forall_{p}\left(\frac{-G M}{r^{2}}-\frac{1}{2} \rho_{s t f} \forall_{p}\frac{d \bold{v}}{d t}\right).
\end{eqnarray}
By defining the effective inertial mass, \(m=\left(3/2\right) \rho_{s t f} \forall_{p}\), we have
\begin{equation}
    F=m\left(\frac{d \bold{v}}{d t}\right)=m\left(\frac{-G M}{r^{2}}\right).
    \label{eq:Newtonforce}
\end{equation}
The forces obtained in \cref{eq:Newtonforce} correspond with the forces acting on the moving test particle around the static point sink.

\subsubsection{Hydrodynamic forces due to the flow field around the rotary point sink}

Now, we study the forces acting on a test particle in the vicinity of a rotating point sink. Rotation of the point sink produces a vortex field. In \citep{40kambe2010new,41kambe2014fluid}, it was shown that there is a connection between acceleration and vorticity fields in an inviscid and compressible fluid flow. In mentioned works, acceleration and vorticity fields for the inviscid compressible flow are defined as
\begin{equation}
\bold{E}=(\bold{u} . \nabla) \bold{u} \ , \quad \boldsymbol{\omega}=\nabla \times \bold{u} \ .
\end{equation}
Which \(\bold{u}\) is the velocity of the fluid, \(\bold{\omega}\) is the vorticity field, \(\bold{E}\) is the acceleration field. The acceleration field is obtained in the \cref{eq:accfield}, and the relation between vorticity and acceleration fields. Suppose that the angular velocity of the rotating sink is very small so that the specific angular momentum approaches zero \((a \rightarrow 0)\). In this case, vorticity and velocity fields  of STF around a rotating point sink, respectively \(\bold{\omega}_{s t f}\) and \(\bold{u}_{s t f}\), will satisfy the below equations.\citep{40kambe2010new}
\begin{align}
    &\nabla . \boldsymbol{\omega}_{s t f}=0, \\
    &\nabla \cdot \frac{D \bold{u}_{s t f}}{D t}=M, \\
    &\nabla \times \frac{D \bold{u}_{s t f}}{D t}+\partial_{t} \boldsymbol{\omega}_{s t f}=0, \\
    &J=c_{s}^{2} \nabla \times \boldsymbol{\omega}_{s t f}-\partial_{t}\left(\frac{D \bold{u}_{s t f}}{D t}\right).
\end{align}
These equations are analogous to the Maxwell equations and its other analogous gravitoelectromagnetic equations. The vortex field around the rotating point sink obtains from the above equations as
\begin{equation}
    \boldsymbol{\omega}_{s t f}=\frac{ G I}{2 c_{s}^{2} r^{3}}\left(\bold{\Omega}-{3(\bold{\Omega} \cdot \hat{r})\hat{r}}\right).
    \label{eq:vorfield}
\end{equation}

To calculate the hydrodynamic forces applied to the test particle around the rotating point sink, we have to place \cref{eq:acceleration,eq:accfield,eq:vorfield}, in \cref{eq:totalforce}. If we assume \(C_{l}=6\) for the STF flow, we will have
\begin{eqnarray}
    \hspace{-1cm}F=&&\left(\frac{3}{2}\right) \rho_{f s t} \forall_{p}\left(\frac{d \bold{v}}{d t}\right)=\left(\frac{3}{2}\right) \rho_{s t f} \forall_{p}\left(\frac{-G M}{r^{2}}\right)\nonumber\\
    &&+4\left(\frac{3}{2}\right) \rho_{c} \forall\left[\bold{v} \times\left(\frac{ G I}{2 c_{s}^{2} r^{3}}(\bold{\Omega}-{3(\bold{\Omega} \cdot \hat{r})\hat{r}})\right)\right].
\end{eqnarray}
Again, we use the effective inertial mass definition, \(m=\left(3/2\right) \rho_{s t f} \forall_{p}\) to reach the following equation: 
\begin{equation}
    F=m\left(\frac{d \bold{v}}{d t}\right)= m\bold{E}+4 m(\bold{v} \times \boldsymbol{\omega}_{s t f}).
    \label{eq:quasiLorentz}
\end{equation}
Here, \(\boldsymbol{\omega}_{s t f}\) corresponds with the GEM field around the rotating mass, and \cref{eq:quasiLorentz} corresponds with the quasi-Lorentz force applied to the NASA probe satellite mentioned in \cref{eq:lorentzforce}. It can be said that the applied forces to the test particle located in the STF around the rotating mass are the same as the forces that are applied to the NASA probe satellite.
\section{Discussion\label{sec:discussion}}
In this paper, we introduced an innovative analogy between fluid dynamics and phenomenological aspects of general relativity, such as black hole physics and gravitoelectromagnetic (GEM). In this analogy, mass is analogous to a point sink in a compressible superfluid we call space-time fluid (STF). Black holes are also modeled as extreme point sinks, around which there are surfaces where the velocity of the fluid reaches the speed of sound. We investigated the hydrodynamic forces applied to a test particle floating in the STF flow using the two-phase flow equations. Also, the acceleration and vorticity fields of the STF flow in the vicinity of the point sink correspond to the gravitational and the GEM fields around the rotary mass, respectively. This analogy could increase our intuitive understanding of the nature of gravity and inertia and provides us with the privilege of acquiring and using powerful fluid dynamics tools such as CFD for modeling and research on gravitational phenomena such as dark matter and Dark energy. According to our analogy, in this section, we will present the foundational discussions about the Mach's principle, the equivalence principle of inertial and gravitational masses, and the lack of information transfer from the event horizon.
\subsection*{Equivalence principle and preferred frame}
The equivalence between the inertial and the gravitational masses is one of Einstein's hypotheses for his general theory of relativity. However, in the presented analogy, according to \cref{eq:Newtonforce}, the inertial and the gravitational masses are proven to be the same. There is no difference in whether the particle takes to be at rest and the STF has acceleration or the STF is stationary, and the particle accelerates with respect to it. We get the same results in both cases, but The latter case is related to the free fall frame while the first case is related to the accelerator frame. Also, according to \cref{eq:accfield}, the acceleration field of the STF flow in the vicinity of the point sink matches with the obtained gravitational field. 
\begin{figure}
    \centering
    \includegraphics[width = 8.6cm]{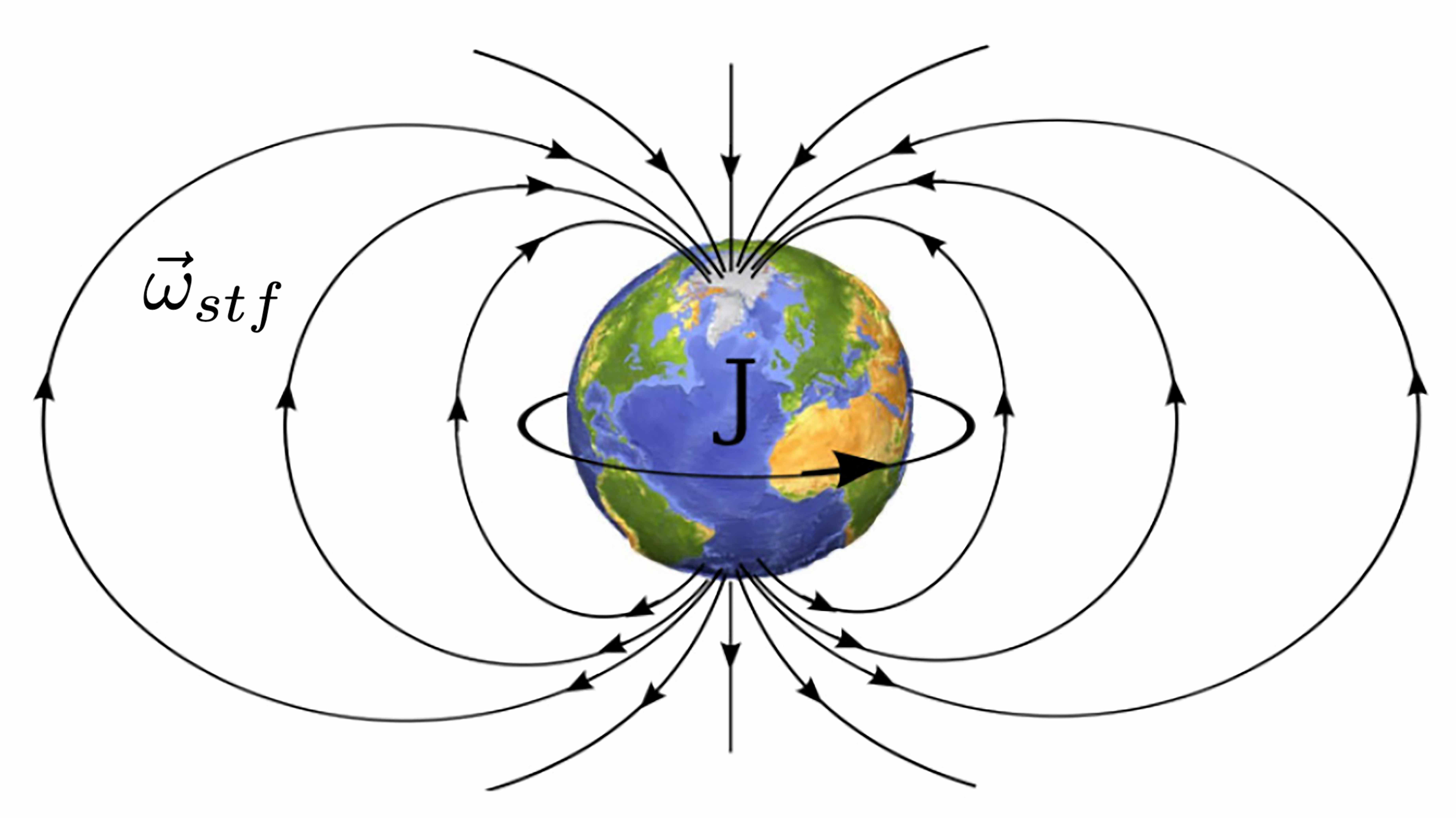}
    \caption{STF vorticity field around a rotating point sink \citep{39gravitomagneticfield}}
    \label{fig:4}
\end{figure}
\subsection*{Mach's principle}
The origin and nature of the force of inertia have always been an issue for physicists. Newton postulates the absolute space-time to answer this question. Regardless, Mach believed in the contextuality in this matter. Mach's principle says that inertia is due to the distribution of the whole objects in the universe. However, Einstein took a mediocre approach in which dynamic space-time and the distribution of objects together determine the inertial frame locally. In this research, the STF flow determines the local reference frame. In this analogy, the relative acceleration between the STF flow and the objects causes the production of a hydrodynamic force that corresponds with the inertial force acting on the accelerating mass.
\subsection*{Information discontinuity in fluid and black hole}
There are unique features on the event horizon and ergosurface around stationary and rotating black holes. For example, light cannot leave the surface of the black hole event horizon. Therefore, the information inside the black hole cannot leave the event horizon surface. In the science of fluid mechanics, there is a discontinuity of information at sonic surfaces because the flow velocity at those surfaces is equal to the velocity of the propagation of acoustic waves. According to recent research on acoustic metrics, the sonic surfaces created inside the converging nozzle correspond with the event horizon surfaces of black holes. Based on this analogy, we calculated the sonic surfaces around the strong point sink that corresponds with the event horizon and the ergosurface around the black holes. This analogy creates an excellent intuitive understanding of the event horizon and ergosphere for us. Also, the properties of the event horizon and the ergosurface around the black hole could be explored.
\section{Conclusions}
This study attempted to investigate the analogy between fluid dynamics and GR. With the help of this analogy, we were able to calculate the relation between the time dilation and the STF flow field. It is shown that the sonic surfaces created around the stationary and rotating point sink correspond with the event horizon and ergosurface around the stationary and rotating black hole. The hydrodynamic forces entered into a test particle placed inside the STF flow field were calculated to validate the obtained flow field equations. The principle of equivalence was discussed due to results that show the acceleration of the fluid flow around the point sink is the same as the gravitational acceleration. Finally, the Mach's principle was discussed, assuming the STF flow as a non-rigid reference. One of the results of this analogy is an increase in the intuitive understanding of gravity and black holes. In addition, we can use the CFD tools for dark energy and dark matter problems and get a clearer picture of the structure inside the black holes.\citep{1klein1973development}
\begin{acknowledgments}
    We sincerely thank Y. Jafari, S. Alapour, M. S. Nomani, H. Fazeli, H. Afshar, H. Niasari, S. Shams, N. Riazi, S. Shahsavar, R. Ahmadvand, K. Goharian, F. Rahmani, M. A. Rafsanjani and Ghadir Jafari for their guidance and assistance.
\end{acknowledgments}

\bibliography{bibliography}

\end{document}